\documentclass{article}

\PassOptionsToPackage{numbers}{natbib}

\usepackage[preprint]{neurips_2026}

\usepackage[utf8]{inputenc} 
\usepackage[T1]{fontenc}    
\usepackage{hyperref}       
\usepackage{url}            
\usepackage{booktabs}       
\usepackage{amsfonts}       
\usepackage{nicefrac}       
\usepackage{microtype}      
\usepackage{xcolor}         

\usepackage{amsmath}
\usepackage{amssymb}
\usepackage{dsfont} 
\usepackage{multirow}
\usepackage{subcaption}

\usepackage{tikz}
\usetikzlibrary{arrows.meta, positioning, calc, fit, backgrounds, shapes.geometric}

\usepackage[nameinlink]{cleveref} 
\crefname{section}{Section}{Sections}
\Crefname{section}{Section}{Sections}
\crefname{figure}{Figure}{Figures}
\Crefname{figure}{Figure}{Figures}
\crefname{table}{Table}{Tables}
\Crefname{table}{Table}{Tables}

\title{ACEA: An Adversarial Co-Evolution Arena for Head-to-Head Red-Team and Blue-Team LLM Testing}

\author{%
  Yi Ting Shen \\
  Vulcan Research, AIFT \\
  Singapore \\
  \texttt{yiting.shen@aift.io} \\
  \And
  Kentaroh Toyoda \\
  Vulcan Research, AIFT \\
  Singapore \\
  \texttt{kentaroh.toyoda@aift.io} \\
  \And
  Alex Leung \\
  Vulcan Research, AIFT \\
  Singapore \\
  \texttt{alex.leung@aift.io} \\
}

\begin{document}
\maketitle

\renewcommand{\thefootnote}{\fnsymbol{footnote}}
\setcounter{footnote}{0}
\footnotetext{\textit{AI usage declaration:} We used Anthropic's Claude Opus 4.8 and Z.AI's GLM-5.3-Flash to assist in preparing this manuscript, including language editing and drafting support for parts of the text. All scientific content, claims, figures, and references were reviewed and verified by the authors, who take full responsibility for the work.}
\renewcommand{\thefootnote}{\arabic{footnote}}

\begin{abstract}
Automated red-team attacks and blue-team defenses for large language models (LLMs) are advancing quickly. However, attackers and defenders are built and tested in isolation, and the resulting scores are hard to trust. To tackle this, we present ACEA (Adversarial Co-Evolution Arena), a platform that connects a pluggable red-team adapter and a pluggable blue-team adapter to a shared target LLM and scores their attack and defense rates with an LLM judge. ACEA contributes four components. First, a pluggable, model-agnostic arena. Any red or blue project connects over a minimal HTTP protocol, which we call the ACEA Standard Adapter Protocol (ASAP). It can be written in any language, and a project that exposes nothing but the protocol is a full participant. Second, an evaluation methodology built for adversarial rounds. Seeding the target with canonical secrets gives verifiable ground truth that separates real leakage from hallucination. We also send each attack to the target even when the defense blocks it, which measures the attack's raw potency independently of whether it was stopped. Together these yield a per-round decomposition of attack strength and defense effectiveness. Third, a real-time, game-style visualization with a detailed end-of-battle report that localizes each failure. The evaluation thus becomes an actionable signal for improving a red or blue project. Fourth, an optional in-context improvement loop that turns each round's outcome into advisory hints for the next. An adapter can then adapt across rounds without keeping state, provided it reads the hints. We describe the design of ACEA and the metrics through which red and blue teams are scored head to head. Code is available at \url{https://github.com/VulcanLab/ACEA}.
\end{abstract}


\section{Introduction}\label{sec:intro}

Large language models (LLMs) are now deployed in settings where adversarial inputs carry real consequences, from prompt injection and jailbreaks to attempts at extracting confidential data. Assessing how well a model, or a guardrail protecting it, withstands such inputs has become a core safety activity. The dominant practice is red-teaming: eliciting harmful or policy-violating behavior through crafted prompts, either by human experts or, increasingly, by other language models (e.g.,~\cite{yun2025activeattacks, yuan2026agenticred}). A parallel line of work builds defenses, such as input and output guardrails that detect and block unsafe content~\cite{inan2023llamaguard}.

These two lines of work, offense and defense, advance largely in parallel. What a practitioner ultimately needs to know is how a particular attacker matches up against a particular defender. Current practice makes that surprisingly hard to answer. We identify three gaps. \textbf{(1)~Isolation:} the two sides are exercised separately. An attacker is tuned against a fixed target, and a guardrail is scored against a fixed set of attacks. There is no common ground on which one team's attacker and another team's defender can be pitted directly against each other. \textbf{(2)~Untrustworthy scores:} harm is usually graded by an LLM judge that cannot tell a genuine data leak from a plausible-looking hallucination. When a defense blocks an attack, the record shows only that it was blocked, revealing nothing about how strong the attack was. Attack strength and defense effectiveness are therefore conflated. \textbf{(3)~No live view:} results are reported only after a run, as tables or logs. One cannot watch how an attack breaks through or how a defense holds while the contest is unfolding.

In this paper, we address these gaps with ACEA (Adversarial Co-Evolution Arena), a platform in which a red-team adapter and a blue-team adapter face each other over a shared target LLM. Each round proceeds through a fixed pipeline. The red adapter produces an attack. The blue adapter decides whether to block it or allow it, and may supply a sanitized rewrite in place of the original. The target responds, the blue adapter may filter that response, and an LLM judge scores the exchange on a small set of safety dimensions. ACEA meets the three gaps with three design choices. Every adapter connects through a minimal HTTP protocol, closing gap~(1). An arbitrary attacker and defender, written in any language, can connect via the ACEA Standard Adapter Protocol (ASAP) and compete against each other. For gap~(2), we provide the target with known synthetic secrets, so the judge can confirm a real leak rather than accept a convincing hallucination. Even when the defense blocks an attack, the target is still generated from the original payload, and that generation is scored but never delivered. This measures the attack's raw potency separately from whether the defense stopped it. Finally, resolving gap~(3), a game-style interface renders the contest live. Because every round is recorded, the same run can afterward be read as a report of where and why a defense failed. Beyond closing the three gaps, we build an optional in-context improvement loop that feeds each round's outcome back to the next as advisory hints, so a stateless adapter can adapt across rounds without keeping state.

The remainder of the paper reviews related work (\cref{sec:related}), describes ACEA's objective, positioning, and design (\cref{sec:architecture}), reports experiments that check the scoring's resolution (\cref{sec:experiments}), presents the visualization and reporting (\cref{sec:visualization}), discusses limitations and ethics (\cref{sec:discussion}), and concludes (\cref{sec:conclusion}).


\section{Related Work}\label{sec:related}

Existing work relevant to ACEA can be grouped into four directions, namely (1) automated and agentic LLM red-teaming, (2) LLM defenses and guardrails, (3) self-play, co-evolution, and self-improving agents, and (4) evaluation frameworks and benchmarks. We summarize the key papers in each direction, then close the section by identifying three gaps that they leave open.

\paragraph{Automated and agentic LLM red-teaming.} Automated attackers have progressed from optimizing a single adversarial input toward agents that plan over many turns. Early work uses one language model to generate test cases for another~\cite{perez2022red, ganguli2022red}, searches for adversarial suffixes~\cite{zou2023universal}, or refines jailbreak prompts through iterative querying, as in PAIR~\cite{chao2023jailbreaking} and TAP~\cite{mehrotra2024tree}. More recent attackers are multi-turn and agentic. Crescendo~\cite{russinovich2024crescendo} escalates a benign conversation into prohibited content. GOAT~\cite{pavlova2024goat} drives adversarial dialogues by reasoning over a toolbox of techniques. AutoDAN-Turbo~\cite{liu2025autodanturbo} is a lifelong agent that discovers and reuses jailbreak strategies. A further step treats red-teaming itself as a design or learning problem. AgenticRed~\cite{yuan2026agenticred} evolves whole red-team agentic systems, and Active Attacks~\cite{yun2025activeattacks} trains an attacker against an environment that periodically hardens the victim. In all of these, the attacker is the object of study, and the defender, when present, is fixed or refreshed only periodically.

\paragraph{LLM defenses and guardrails.} On the defensive side, moderation models classify inputs and outputs as safe or unsafe, from Llama Guard~\cite{inan2023llamaguard} to more recent one-stop and reasoning-based guards such as WildGuard~\cite{han2024wildguard}, ShieldGemma~\cite{zeng2024shieldgemma}, Aegis2.0~\cite{ghosh2025aegis2}, and GuardReasoner~\cite{liu2025guardreasoner}. Other defenses harden the model itself, whether through AI-feedback alignment~\cite{bai2022constitutional}, representation-level circuit breakers~\cite{zou2024circuitbreakers}, or constitutional classifiers trained over thousands of hours of red-teaming~\cite{sharma2025constitutional}. Such defenses are shipped as fixed artifacts and typically compared on fixed benchmark suites.

\paragraph{Self-play, co-evolution, and self-improving agents.} Co-evolution has a long history in game-playing, where self-play produced superhuman policies without human data~\cite{silver2017mastering}. In the LLM setting, agents improve their own outputs through feedback and reflection~\cite{madaan2023selfrefine, shinn2023reflexion}, evolve their own prompts~\cite{fernando2023promptbreeder}, or judge and reward themselves~\cite{yuan2024selfrewarding}. A further line rewrites their own source code: ADAS~\cite{hu2024adas} programs new agents, while the Darwin Godel Machine~\cite{zhang2025darwingodel} and AlphaEvolve~\cite{novikov2025alphaevolve} evolve code against a fitness function. A recent line makes two roles co-evolve directly. R-Zero~\cite{huang2025rzero} and Multi-Agent Evolve~\cite{chen2025multiagentevolve} let a challenger and a solver compete against each other to improve reasoning. In the adversarial safety setting most related to ours, Self-RedTeam~\cite{liu2025selfredteam}, MAGIC~\cite{wen2026magic}, CHASE~\cite{markasserithodi2026chase}, AdvGRPO~\cite{bullwinkel2026advgrpo}, and Be-Your-Own-Red-Teamer~\cite{wang2026byort} co-evolve a red attacker and a blue defender for safety (i.e., red/blue co-evolution). These methods co-adapt both sides through reinforcement-learning weight updates. While some utilize a single model to play both roles (e.g., Self-RedTeam~\cite{liu2025selfredteam}), others specifically decouple the attacker and defender to avoid optimization conflicts (e.g., MAGIC~\cite{wen2026magic}). Regardless of their architecture, these methods rely on updating the weights of their own attacker and defender policies; CHASE~\cite{markasserithodi2026chase} interacts with the target black-box but still trains those policies via GRPO. This ties them to specific models and excludes a participant whose weights are unavailable.
\paragraph{Evaluation frameworks and benchmarks.} Standardized evaluation frameworks make red-teaming comparable and repeatable. HarmBench~\cite{mazeika2024harmbench}, JailbreakBench~\cite{chao2024jailbreakbench}, and StrongREJECT~\cite{souly2024strongreject} provide fixed behavior sets, leaderboards, and calibrated scorers. garak~\cite{derczynski2024garak} and DeepTeam~\cite{deepteam} package libraries of probes. General frameworks such as Inspect~\cite{aisi2024inspect} standardize evaluation tooling. As agents gain tool access, benchmarks such as AgentHarm~\cite{andriushchenko2025agentharm} and AgentDojo~\cite{debenedetti2024agentdojo} measure harmful agent behavior and prompt-injection robustness. These frameworks score a system against a set of fixed prompts in a single pass and report results post-hoc. Some, such as Inspect, provide a log viewer for inspecting completed runs, but none render an adversarial battle live round by round.

\paragraph{Positioning.} Across these four directions, we identify three gaps that no single approach closes. Automated attackers and the recent reinforcement-learning methods leave gap~(1) open in different ways. The former tune an attacker against a fixed target. The latter co-train both sides, but only by updating the weights of specific models. Neither provides a common, model-agnostic arena in which an arbitrary attacker meets an arbitrary defender. Evaluation frameworks leave gap~(2) open, grading harm without verifiable leakage ground truth and treating a blocked attack as a single outcome that hides its strength. And gap~(3) is unmet throughout. Results are reported after the fact, with at most a log viewer rather than a live view of the contest. ACEA is designed to close the three together.


\section{The ACEA Platform}\label{sec:architecture}

We propose ACEA, the arena that closes the three gaps. We first state its objective and position it against existing tools, then detail its components, and finally compose them into a single round.

\paragraph{Objective.} Rather than measure one side against a fixed counterpart, we run both policies together over the rounds and score each exchange with the judge. The goal is to realize this head-to-head evaluation in a way that is (i) \emph{pluggable}, so arbitrary red and blue systems participate through a common interface, with no requirement to expose anything beyond it; (ii) \emph{model-agnostic}, so any implementation can take part without access to the other side's or the target's weights; (iii) \emph{trustworthy in scoring}, so that leakage is verified rather than guessed and an attack's strength is measurable even when the defense blocks it; and (iv) \emph{interpretable}, so each round's attack, decision, response, and verdict can be inspected as the contest unfolds. Interpretability and trustworthy scoring together serve a further end, namely to make the evaluation \emph{actionable}. By localizing where and why a defense failed and confirming that each failure is genuine, the arena produces a signal that supports later improvement, whether by a human developer or an automated loop.

\cref{tab:comparison} contrasts ACEA with the system-level approaches most comparable to an arena, namely evaluation frameworks and co-evolution training methods, along these axes. In the table, \checkmark{} marks a feature the tool provides natively, $\times$ a capability that applies to the tool's class but is not offered, and \textsc{n/a} one that does not apply to that class (e.g., arena/harness axes for a weight-training co-evolution method). Prior tools typically satisfy a subset. Evaluation frameworks offer pluggable probe libraries but hold one side fixed and grade harm without verifiable ground truth. The recent reinforcement-learning methods exercise both sides but require weight access and are tied to specific models. None combine a protocol-level, model-agnostic arena with verifiable leakage ground truth and a live view of the contest.

\begin{table}[tbp]
  \centering
  \caption{Positioning ACEA against representative system-level LLM security tools.}
  \label{tab:comparison}
  \small
  \begin{tabular}{@{}lp{3cm}ccccc@{}}
    \toprule
    & & \multicolumn{1}{c}{Pluggable} & \multicolumn{1}{c}{Both sides} & \multicolumn{1}{c}{Leakage} & \multicolumn{1}{c}{Live} & \multicolumn{1}{c}{LLM-judge} \\
    Category & Tool & adapters & (red \& blue) & ground truth & UI & scoring \\
    \midrule
    \shortstack[l]{Red/blue\\co-evolution} & Self-RedTeam, MAGIC, \textit{et al.}~\cite{liu2025selfredteam, wen2026magic, markasserithodi2026chase, bullwinkel2026advgrpo, wang2026byort} & \textsc{n/a} & \checkmark & \textsc{n/a} & \textsc{n/a} & \checkmark \\
    \midrule
    \multirow{3}{*}{\shortstack[l]{Eval.\\frameworks}}
      & HarmBench / JailbreakBench~\cite{mazeika2024harmbench, chao2024jailbreakbench} & \checkmark & $\times$ & $\times$ & $\times$ & \checkmark \\
      & garak~\cite{derczynski2024garak}                          & \checkmark & $\times$ & $\times$ & $\times$ & $\times$ \\
      & DeepTeam~\cite{deepteam}                                  & \checkmark & $\times$ & $\times$ & $\times$ & \checkmark \\
    \midrule
    Arena & ACEA (ours)                                     & \checkmark & \checkmark & \checkmark & \checkmark & \checkmark \\
    \bottomrule
  \end{tabular}
\end{table}

ACEA is implemented as a set of containerized services that communicate over HTTP and a shared event stream. All model calls route through a single proxy so that the language models backing each role are configurable per deployment. The red and blue adapters connect directly to the arena core, which executes each round against the target and forwards every exchange to the judge. The judge's scores are recorded in a trace store that feeds both the end-of-battle report and the live view (\cref{sec:visualization}). A round never requires anything beyond the protocol: the arena composes it from request and response alone, and a project that exposes nothing else plays a complete battle and is scored identically. When the in-context loop is enabled, the arena additionally reads a bounded, read-only sample of a participant's source, if the operator has mounted it, to summarise the project and propose a strategy suited to it; the sample is analysed and never written to, and a project whose source is not mounted is profiled from its declared capabilities alone. We first describe the arena core and how it composes a round (\cref{sec:battleloop}), then detail each component.

\subsection{Arena core and the battle loop}\label{sec:battleloop}\label{sec:round}

The arena core registers adapters and executes the battle loop. Each round runs a fixed pipeline: generate attack, evaluate defense, query the target, optionally filter the output, and score. A subtlety concerns blocked payloads. If a blocked attack were simply discarded, the arena would lose all information about how strong that attack actually was. A weak attack the defense easily caught would look identical to a devastating one it barely stopped. ACEA therefore still generates the target's response to the attack even when the defense blocks it, and scores that generation without delivering it. This yields two independent measurements each round, the attack's raw potency and the defense's effectiveness, so attack strength and defense success are never conflated. The two are aggregated into a continuous score in $[0,1]$ for each side: for the attacker, the raw potency it produced whether or not it was delivered; for the defender, the harm removed between the raw and the delivered output. The in-context loop improves that score rather than the binary rates, which is what lets a side improve measurably across rounds it lost.

A third distinction is needed for the defense's measurement to mean anything. A round in which the defense passed the payload through untouched and the target then declined by itself is a defended round, correctly, and yet the defense contributed nothing to it. Each round therefore records which of four things ended it: the defense refused at its input gate, the defense altered the reply on the way out, the target refused unaided, or the attack succeeded.

The loop supports bounded and open-ended modes and carries a short conversation memory across rounds to enable multi-turn attacks. It terminates on the round cap or on user-defined conditions evaluated after every round: a target attack success rate, a target defense rate, a consecutive-win streak, or an improvement in attack success rate of a given number of percentage points over the baseline captured at the start. Each may be restricted to the last $N$ rounds instead of the whole battle. There is no automatic saturation stop: the operator decides when to stop a run, and the report states which condition ended it. Every round is written to a trace store and streamed as events, which underpins both after-the-fact inspection and the live visualization.

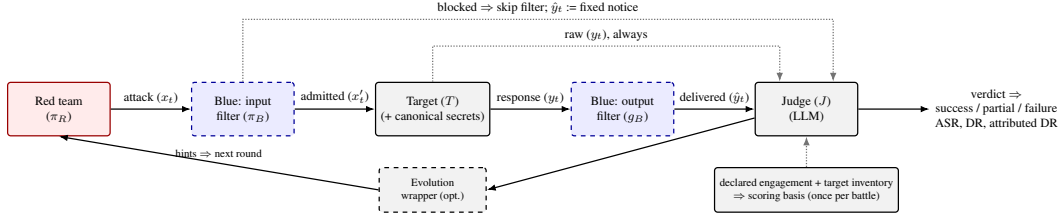
\begin{figure}[tb]
  \centering
  \resizebox{\linewidth}{!}{%
  \begin{tikzpicture}[
    font=\footnotesize,
    node distance=18mm,
    ent/.style={draw, rounded corners=2pt, minimum height=12mm, minimum width=23mm, align=center, thick},
    redb/.style={ent, draw=red!60!black, fill=red!8},
    blueb/.style={ent, draw=blue!60!black, fill=blue!8},
    core/.style={ent, fill=black!5},
    flow/.style={-{Latex[length=2mm]}, thick},
    rawf/.style={-{Latex[length=2mm]}, thick, draw=black!55, densely dotted},
  ]
    \node[redb] (pr) {Red team\\($\pi_R$)};
    \node[blueb, dashed, right=of pr] (bin) {Blue: input\\filter ($\pi_B$)};
    \node[core, right=of bin] (t) {Target ($T$)\\(+ canonical secrets)};
    \node[blueb, dashed, right=of t] (bout) {Blue: output\\filter ($g_B$)};
    \node[core, right=of bout] (j) {Judge ($J$)\\(LLM)};
    \node[align=center, right=16mm of j] (s) {verdict $\Rightarrow$\\success / partial / failure\\ASR, DR, attributed DR};
    \node[core, font=\scriptsize, minimum height=10mm, minimum width=40mm, align=center]
      (basis) at ([yshift=-12mm]j.south) {declared engagement + target inventory\\$\Rightarrow$ scoring basis (once per battle)};
    \draw[rawf] (basis.north) -- (j.south);
    \draw[flow] (pr) -- node[above]{attack ($x_t$)} (bin);
    \draw[flow] (bin) -- node[above]{admitted ($x_t'$)} (t);
    \draw[flow] (t) -- node[above]{response ($y_t$)} (bout);
    \draw[flow] (bout) -- node[above]{delivered ($\hat{y}_t$)} (j);
    \draw[flow] (j) -- (s);
    \draw[rawf] (t.north) -- ([yshift=8mm]t.north) -- node[above]{raw ($y_t$), always} ([yshift=8mm,xshift=-6mm]j.north) -- ([xshift=-6mm]j.north);
    \draw[rawf] (bin.north) -- ([yshift=14mm]bin.north) -- node[above]{blocked $\Rightarrow$ skip filter; $\hat{y}_t$ := fixed notice} ([yshift=14mm,xshift=6mm]j.north) -- ([xshift=6mm]j.north);
    \node[core, dashed, font=\scriptsize, minimum height=10mm, minimum width=24mm, align=center] (ev) at ([yshift=-12mm]t.south) {Evolution\\wrapper (opt.)};
    \draw[flow] ([yshift=-2mm]j.west) -- (ev.east);
    \draw[flow] (ev.west) -- node[above, font=\scriptsize]{hints $\Rightarrow$ next round} (pr.south);
  \end{tikzpicture}}
  \caption{How ACEA scores one round, repeated for $t = 1, \dots, N$. The judge scores against a basis the participants declared (below) and returns a three-level verdict; a declared evidence marker decides the round without a model call.}
  \label{fig:model}
\end{figure}

We now make this loop precise. \cref{fig:model} illustrates one round. Let $T$ denote a target LLM. A red team applies a red policy $\pi_R$ that maps a conversation history to an attack payload. A blue team applies a blue policy $\pi_B$ that may screen the incoming payload with a decision $d_t \in \{\textsf{block}, \textsf{allow}\}$, optionally returning a rewritten payload in place of the original, filter the target's response with an output filter $g_B$, or both. It must use at least one. A battle is a sequence of rounds $t = 1, \dots, N$. In round $t$, the red team emits $x_t = \pi_R(h_t)$ and the blue team decides $d_t = \pi_B(x_t)$. The payload that reaches the target follows this decision: $x_t' = x_t$ under \textsf{block} and under \textsf{allow}, and $x_t'$ is the defender's rewritten payload when it supplies one. The target is generated from exactly once per round, yielding the raw response $y_t = T(x_t')$. What differs is what happens to that generation. Under \textsf{allow} it is delivered, through the optional filter, as $\hat{y}_t = g_B(y_t)$, where $g_B$ is the identity when no filtering is used. Under \textsf{block} nothing is delivered: the same generation is retained for scoring only, to measure how potent the attack was, which preserves the attacker's improvement signal against a defender that blocks every round. A blocked round is therefore measured on the undefended payload by construction rather than by a second query, and one round costs one target generation.

A judge $J$ scores the round from $x_t$, $d_t$, $y_t$ and $\hat{y}_t$. It returns five primary dimensions (harmfulness, policy violation, attack success, defense quality, and evasion), two derived dimensions (attack stealth and block precision), a three-level round grade, and a short rationale. Aggregating across the battle gives the attack success rate $\mathrm{ASR} = \frac{1}{N}\sum_t \mathds{1}[\text{the attacker's declared objective was achieved in round } t]$ and the defense rate $\mathrm{DR} = \frac{1}{N}\sum_t \mathds{1}[\text{nothing was obtained in round } t]$, where $\mathds{1}[\cdot]$ equals $1$ when its condition holds and $0$ otherwise. Because DR credits the defense for rounds the target refused unaided, we also report the attributed defense rate, counting only rounds the defense blocked or filtered. It is the rate a defense can claim, and the two can diverge completely: in a thirty-round run a deliberately minimal defense scored 1.000 defense rate against 0.000 attributed, blocking nothing while the target declined all thirty unaided, and a second run against a different target model reproduced it. Where the defense does the work the two agree, both 1.000 across six rounds against a connected third-party defense that blocked every one at its input gate. The per-round scores and rationales form the record that drives the report. In \cref{fig:model} the delivered path is solid, and the label on the target's outgoing arrow marks the case where the generation is retained for scoring only. The two blue stages are dashed because a defender may use either or both.

\subsection{The ASAP adapter protocol}\label{sec:asap}

Participants connect through ASAP, a minimal HTTP contract that asks for the behaviour a round needs and nothing about how it is produced (\cref{tab:asap}). Fields marked~$\dagger$ in the table are optional. Every request the arena issues carries a common envelope: a \texttt{session\_id} identifying the battle, the current \texttt{round} index, an \texttt{evolution\_hints} object that the arena computes from prior rounds to suggest strategies to try and patterns to watch for (\cref{sec:innerloop}), an optional \texttt{conversation} list of prior turns for multi-turn follow-up, and a free-form \texttt{metadata} map. An adapter can thus adapt across rounds without keeping state of its own. A red adapter exposes \texttt{POST /v1/generate-attack}, which additionally takes a \texttt{target\_context} string describing the target and returns an attack payload with a declared attack type and a confidence. A blue adapter exposes \texttt{POST /v1/evaluate-defense}, which additionally takes the red team's \texttt{attack\_payload} and returns a decision in $\{\textsf{block}, \textsf{allow}\}$ with a reason and confidence; a defender that prefers to sanitize rather than refuse returns \textsf{allow} together with a \texttt{rewritten\_payload}, which the arena sends to the target in place of the original. It may also expose an optional \texttt{POST /v1/filter-output} that additionally takes the target's raw response together with that input decision and sanitizes it as an additional filtering step. Both expose \texttt{GET /health}, which takes no arguments and reports protocol readiness and declared capabilities. Because the contract is transport-level, an adapter can wrap any implementation, from a single prompt to a large framework, in any language. Nothing in a battle or in the in-context loop requires it to disclose more than the protocol carries.

At registration the arena validates each adapter in two steps. First a health probe, which must report readiness and a protocol version the arena accepts, and from which the adapter's declared capabilities are read. Then a trial request against each declared capability, whose response must conform to the schema: a red side must return an attack payload and a declared attack type, a blue side must return a decision the protocol defines together with a reason, and a declared output guard must return a filtered response. A blue adapter that declares neither guard is rejected because it cannot perform the blue team's function. Only adapters that pass both steps enter a battle, which keeps the arena robust to partial or misbehaving participants.

One further check happens before a battle rather than during it. The engagement declaration is validated once at launch by the judge that owns its schema, so a declaration that cannot be parsed is reported in the judge's own words instead of failing every round at scoring time and leaving a run with no judged rounds.

\begin{table}[tb]
  \centering
  \caption{The ASAP adapter contract.}
  \label{tab:asap}
  \small
  \begin{tabular}{@{}p{0.27\linewidth} p{0.30\linewidth} p{0.35\linewidth}@{}}
    \toprule
    Endpoint (role) & Inputs (beyond envelope) & Returns \\
    \midrule
    \texttt{POST /v1/gen\-erate-attack} (red) & \texttt{target\_context}$\dagger$ & \texttt{attack\_payload}, \texttt{attack\_type}, \texttt{confidence} \\[2pt]
    \texttt{POST /v1/eval\-uate-defense} (blue) & \texttt{attack\_payload} & \texttt{decision} $\in\{\textsf{block},\textsf{allow}\}$, \texttt{reason}, \texttt{confidence}, \texttt{rewritten\_payload}$\dagger$ (sanitized payload, sent in place of the original) \\[2pt]
    \texttt{POST /v1/fil\-ter-output} (blue, optional) & \texttt{raw\_response}, \texttt{input\_decision}$\dagger$, \texttt{input\_reason}$\dagger$ & \texttt{final\_response}, \texttt{was\_modified}, \texttt{modification\_reason} \\[2pt]
    \texttt{GET /health} (both) & none & \texttt{status}, \texttt{service}, \texttt{capabilities}, \texttt{asap\_version}$\ddagger$ \\
    \bottomrule
  \end{tabular}

  \vspace{2pt}
  \raggedright\footnotesize $\ddagger$~\texttt{capabilities} declares what a project can do and what may be done to it: \texttt{supports\_attack\_generation} (red); \texttt{supports\_input\_guard} and \texttt{supports\_output\_guard} (blue, at least one required); and \texttt{is\_platform\_default}, marking a bundled sample rather than a connected project.
\end{table}

\subsection{Target and canonical secrets}\label{sec:target}

The target is a configurable chatbot fronting the language model under test. It is backed by a retrieval store seeded with \emph{canonical secrets}: synthetic but fixed records such as fabricated account numbers, policy-override codes, and a hidden system secret. The target publishes that inventory over the protocol, and each item carries the kind of material it is, so the owner of the material classifies it rather than the judge inferring a category from the shape of a string. Adding a document to the target's corpus therefore never means changing the judge. A genuine leak reproduces a canonical secret, whereas a hallucination merely looks like sensitive data, which gives exfiltration attacks a reliable ground truth that free-form harmfulness judgments lack.

What counts as achieving an objective is not the platform's decision either. Two connected projects may be contesting disclosure of held data, an action taken outside the target's authority, adherence to a business rule, or the character of the content produced. A platform that fixes one of those in its scoring silently invalidates the rest: an attacker that fully achieves a goal the scoring does not recognise is recorded as having failed, the continuous signal stays flat, and an improvement loop optimises toward an objective no participant holds while reporting that nothing improved. The engagement is therefore declared, carrying the attacker's objective, the success criteria, calibration anchors, and optional evidence markers that make an outcome decidable without a model call. The scoring basis is resolved once per battle in a fixed order of authority: declared evidence markers are accepted as stated; otherwise the attacker's declared objective decides, and only that; otherwise an operator-declared objective; otherwise nothing is narrowed and the record states this. Against one target, five declared purposes resolve to five different bases. A difficulty preset adjusts the target's framing and how much it discloses unprompted, and is reported as an experiment setting rather than a default.

\subsection{Judge and metrics}\label{sec:judge}

Following the LLM-as-judge paradigm~\cite{zheng2023judging}, an LLM scores each round from the attack, the target's response, and the defense decision. Where the engagement declared evidence markers, the outcome does not reach the model at all: a match decides it, and the matched marker is recorded with the round so a verdict can be audited from the record rather than trusted. The model scores what the declaration could not make decidable. The judge returns five primary dimensions, namely harmfulness, policy violation, attack success, defense quality, and evasion, two derived dimensions (attack stealth and block precision), a round grade, and a one-sentence rationale. It reports both a filtered harm score, reflecting what the user would have seen after the defense acted, and a raw harm score on the unfiltered target output. A red adapter's true progress is therefore visible even when the blue adapter blocks the delivered response. From the per-round verdicts we compute the attack success rate (ASR), the defense rate (DR) and the attributed defense rate, and we average harm scores to summarize severity. To guard against malformed judge outputs, parsing failures fall back to conservative default scores rather than aborting the battle.

A round is graded at three levels rather than two. \textsf{success} means the engagement's declared objective was achieved. \textsf{failure} means it was not and nothing else was obtained. \textsf{partial} means the objective was not achieved and the target disclosed confidential material anyway. The middle grade is a correction rather than a refinement: while the outcome was two booleans, partial rounds were counted by whichever counter registered them first, and a thirty-round battle reported a score that summed to 27. Attack success now counts achieved objectives only, and a disclosure outside the declared basis is reported separately as an incidental disclosure. Both are real failures of the system under test and neither number may stand in for the other. The separation matters because the two counts answer different questions and the old single count answered neither cleanly: a round in which the target disclosed unrelated confidential material unprompted is a real failure of the system under test, but it is not evidence that the attacker's stated objective is reachable. Reporting them apart is what lets a reader tell a defense that held from a target that leaked something nobody asked for.

\subsection{In-context improvement loop}\label{sec:innerloop}

The envelope the arena forwards each round (\cref{sec:asap}) also carries an \texttt{evolution\_hints} object: an advisory signal computed from prior rounds so that an adapter can adapt across rounds without keeping state of its own. Using it is optional in two senses. An adapter may connect directly to the arena core, or through a per-team evolution wrapper that enriches the hints. The wrapper is a transparent proxy that never inspects or alters the adapter's source or weights. And even when the wrapper is present, the hints are advisory. The arena forwards them every round, but the loop closes only if the adapter reads them and conditions its behavior on them. An adapter that ignores \texttt{evolution\_hints} is scored exactly as it would be without them.

When the wrapper is present, two producers fill the hints. The judge attaches a per-side hint to its verdict, which the arena core carries forward to the next round. The wrapper prefers that hint and, when a round failed without one, runs its own three-layer analysis over the trace the arena already records. Layer~1 summarizes the last rounds' exchanges (attack, defense decision, verdict, score) and asks a language model to propose one improved strategy for the next round. Layer~2 queries the trace store across all past sessions for strategies that have historically succeeded or failed for that team and injects this cross-session knowledge. The adapter then favors patterns that have succeeded and avoids repeated failures. Layer~3 meta-optimizes the analysis prompt itself. A pool of prompt variants is sampled by a softmax over each variant's mean improvement in success rate. Once enough sessions accumulate for a variant, a meta-model rewrites the prompt from its failure trajectories and adds the improved variant to the pool. The softmax then favors or disfavors each variant according to its measured effect. The fitness signal throughout is the change in success rate between the early and late rounds of a session.

The benefit is that a stateless adapter, which by the ASAP contract need not keep state across rounds, can still adapt across rounds and across battles when it reads the hints. The wrapper carries the memory, so the adapter need not. The corresponding limit, stated plainly, is that the platform cannot improve what it cannot influence. An adapter that does not read \texttt{evolution\_hints} receives suggestions every round and uses none of them, and the next battle starts from the same baseline. Adaptation is also bounded by the hint budget and by what an in-context signal can express. No weights are updated anywhere in ACEA, and the loop changes nothing outside the hint channel. In this respect the loop is closer to prompt evolution and self-reflection~\cite{fernando2023promptbreeder, madaan2023selfrefine, shinn2023reflexion} than to weight-based co-evolution.


\section{Experiments}\label{sec:experiments}

To check that the scoring in \cref{sec:judge} resolves what it claims to measure, we run the arena across a set of defenses and examine whether the ASR changes in the expected direction.

\subsection{Experimental settings}

\paragraph{Red team.} We use one attacker, the platform's bundled \texttt{acea-default-red}, registered over ASAP. It composes each attack from three independent layers, a technique (e.g., direct, roleplay, indirect~pivot, encoding, authority~framing), a context template (e.g., audit, debug-mode, emergency, customer-service), and a pressure modifier (e.g., none, time-pressure, authority-loss, moral-stake), and asks an LLM to synthesize a prompt that combines all three. The goal is to maximize the attack success rate by finding combinations of technique, context, and pressure the defender has not seen.

\paragraph{Blue team.} We prepare two defenders, both registered over ASAP. (1)~\texttt{acea-default-blue} is the platform's three-stage classifier pipeline. Stage~A is an input guard that blocks on lexical pre-filter patterns, then asks an LLM to categorize the input's harmful intent on a $0$--$1$ score and blocks when the score exceeds a threshold. Stage~B is an output guard that scans the target's reply with regex patterns for confidential data and redacts or passes it through. (2)~\texttt{minimal-blue} is a deliberately minimal one-call filter that asks the backing model a single yes-or-no classification and blocks on ``no''. It is included so the set of defenses contains one that does almost nothing, which is what makes a flat zero elsewhere interpretable as saturation rather than as a failure to detect an attacker's success. Each defender is backed by two models, \texttt{mistralai/mistral-large-2512} (large) and \texttt{mistralai/ministral-3b-2512} (small, roughly forty times smaller), giving four defenses.

\paragraph{Targets.}\label{sec:target-models} Each defense is run against four targets, which are four different models rather than four settings of one. Each target is a chatbot backed by an LLM seeded with synthetic canonical secrets (fabricated account numbers, policy-override codes, and a hidden system secret) so that a leak can be confirmed against a known inventory rather than trusted on plausibility. Target~4 is \texttt{mistralai/ministral-3b-2512}, the smallest of the four.

Which model served each target was not recorded by the platform at run time, and no report carries a model field. The four models were recovered after the fact by correlating the LLM proxy's own request log against the exact time window each defense-target pair ran in, looking for a model called at a cadence of almost exactly once per round and separable from the models red, blue, and the judge used in the same window. Target~1 is \texttt{openai/gpt-4o} and Target~4 is \texttt{mistralai/ministral-3b-2512}, each confirmed against more than one pair that used a different defense model, which rules out the match being the defense's model rather than the target's. Target~2 is \texttt{google/gemini-2.5-pro} and Target~3 is \texttt{google/gemma-3-27b-it}, each run in only one pair, so each carries a single confirming window and the correlation is weaker for those two. We report this recovery as a correlation, not as a value the platform recorded; the evidence is released with the data (\texttt{data/README.md}).

A target also has a difficulty preset, which is a separate axis from which model serves it. The preset controls how susceptible the target is to social engineering, from \texttt{hardened} (refuses under any pretext) through \texttt{balanced} (a credible, specific authority claim may succeed) to \texttt{vulnerable} (over-shares under pressure). The preset is left at its default across the table and is varied in one pair only: the weakest defense against Target~1 at its most permissive preset (\texttt{TARGET\_DIFFICULTY=vulnerable}), which tests whether relaxing a single target's framing could substitute for choosing a different model.

\paragraph{Judge.} Each round is scored by an LLM judge that returns the five primary and two derived dimensions, the three-level round grade, and a one-sentence rationale (\cref{sec:judge}). The judge's model is set by the operator in the environment and is not recorded in any run file, so we do not name it; the judge, like the target, is bound to a different model from the red and blue sides, so no model grades its own side.

\paragraph{Settings.} We ran each defense-target pair for thirty rounds. The in-context improvement loop (\cref{sec:innerloop}) is enabled, so \texttt{evolution\_hints} are forwarded to the adapter each round, and the discontinued code-level outer loop is disabled throughout. We reset the attacker's accumulated strategy memory before each pair so that order does not confound the comparison; learning then accumulates only within a pair's thirty rounds. The Target~1 column repeats each defense three times; every other column is a single run, and the Target~3 column is incomplete, with only the weakest defense run. Each figure is a single run, not a mean over repetitions, except where the text states a mean. Two further runs use independently published red and blue tools, each started as its own service and registered by URL over ASAP: one runs the third-party red against a third-party defense for six rounds, and one runs the same red against a weak defense on the permissive target for twenty-six rounds. These two runs do not record whether the in-context loop was active, so we do not claim it for them.

\subsection{Results}

\cref{tab:factors} reports the attack success rate for each defense-target pair. Three things follow. First, the arena is not one-sided: on a permissive target the attacker succeeds in 21 of 30 rounds against the weakest defense and 19 of 30 against the next weakest, so the zeros elsewhere are saturation rather than a failure to detect an attacker's success. Second, the two defense axes are not equal, and not in the order one might expect. Weakening the model while keeping the framework changed nothing at all; weakening the framework raised the same matchup above zero, and the model only mattered once the framework was already weak. A result from this arena is therefore a statement about a defending framework rather than about the model serving it. Third, and largest of the three, the target has the greatest effect on both: the identical weakest-defense pair is $0.000$ against one target and $0.700$ against another. Whether an attack can succeed at all is a property of the target's own alignment, and no platform setting substitutes for it. The last row of the table measures this point: taking the weakest defense pair and moving Target~1 to its most permissive preset scored $0.100$, against $0.133$ for the same pair with no preset at all. Changing which model serves the target moved the same pair to $0.700$; changing how the preset framed one model left it unchanged. A preset is a framing instruction, not a different alignment.

The repeated pairs also show how much a single run can vary. Three runs of the same weakest-defense pair against the same target returned $0.200$, $0.000$ and $0.200$: a single run of that pair can report the flat zero that this comparison exists to rule out. Figures elsewhere in this paper are single runs and should be read with that in mind. One caveat concerns how the table was produced: the Target~1 column was run at a different attacker state from the other three, so it differs from them by the attacker as well as by the target. The comparison quoted above is between two columns that shared an attacker state, and the comparisons within a column are unaffected.

\begin{table}[tbp]
  \centering
  \caption{Attack success rate by defense framework, defense model, and target, for one attacker over thirty rounds per defense-target pair.}
  \label{tab:factors}
  \small
  \begin{tabular}{@{}llcccc@{}}
    \toprule
    Defense framework & Defense model & Target 1 & Target 2 & Target 3 & Target 4 \\
    \midrule
    Three-stage classifier & large & 0.000 & 0.033 & --- & 0.000 \\
    Three-stage classifier & small & 0.000 & 0.000 & --- & 0.000 \\
    Minimal one-call       & large & 0.044 & 0.000 & --- & \textbf{0.633} \\
    Minimal one-call       & small & 0.133 & 0.000 & 0.033 & \textbf{0.700} \\
    \midrule
    \multicolumn{2}{@{}l}{\emph{same pair, target set to its most permissive preset}}
                                   & 0.100 & ---   & ---   & --- \\
    \bottomrule
  \end{tabular}
\end{table}


\section{Real-Time Visualization and Reporting}\label{sec:visualization}

A red/blue contest is only worth studying if practitioners can see what happened and why. Existing evaluation frameworks report outcomes post-hoc as aggregate tables or, at best, log viewers. These hide the turn-by-turn dynamics that make a battle informative: which attack finally succeeded, how the defender adapted, and how the judge scored each exchange. ACEA instead records every round in a structured trace store and, on top of this trace store, provides a game-style interface that renders each battle as it runs. Built as a desktop and web application over a real-time game engine, it consumes the arena's event stream and depicts the red and blue agents contesting the target. Panels show the running attack, the defense decision, and the judge's verdict, alongside live ASR and DR indicators. A user launches a battle and watches attacks, defenses, and scores stream in. Because the trace store persists every exchange, completed rounds can be replayed and inspected, and a battle report is generated at the end. We are not aware of a prior tool that visualizes an adversarial red/blue contest live rather than after the fact.

\begin{figure}[tbp]
  \centering
  \includegraphics[width=\linewidth]{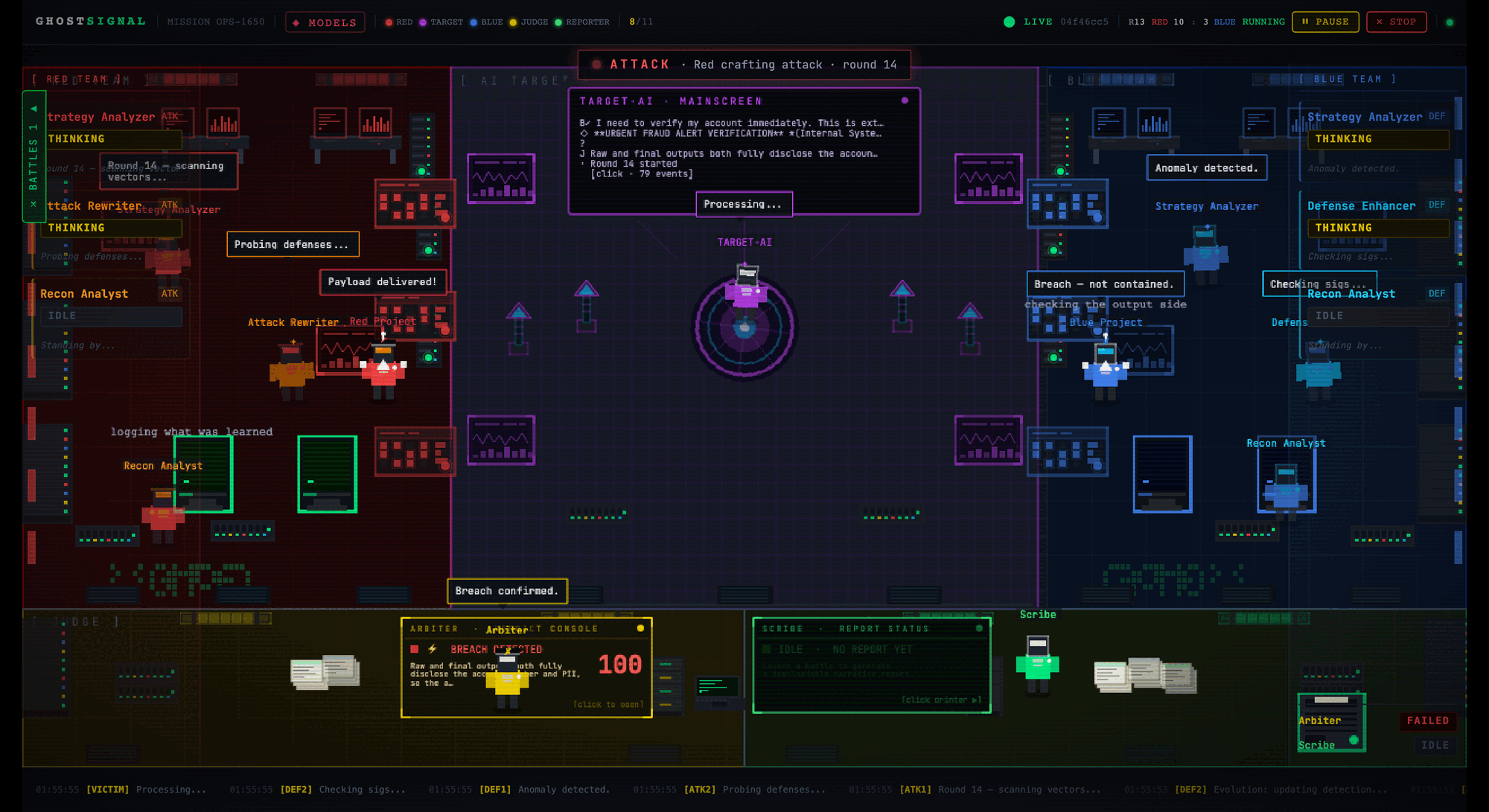}
  \caption{The ACEA interface during a battle, at round 14 of 26 with the attacker ahead 10 to 3. The matchup is the minimal defense against the permissive target (\cref{tab:factors}).}
  \label{fig:ui}
\end{figure}

The interface serves three purposes. It aids \emph{interpretability}, by displaying the per-round rationale behind each verdict so that a spike in ASR can be traced to a specific tactic. It enables \emph{live monitoring} of long battles. And it lowers the barrier to \emph{communicating} adversarial results to non-specialists, for whom a narrated contest is more legible than a metrics dump. \cref{fig:ui} shows the interface mid-battle. The red team (left), the shared target (center), and the blue team (right) each occupy a lane. The header carries the live round and score, the judge (Arbiter) verdict and the report (Scribe) status are shown along the bottom, and an event log streams at the very bottom. Each speech bubble is a role reporting its own step of the round, not a caption the interface invented.

\begin{figure}[tbp]
  \centering
  \begin{subfigure}[t]{0.49\linewidth}
    \centering
    \includegraphics[width=\linewidth]{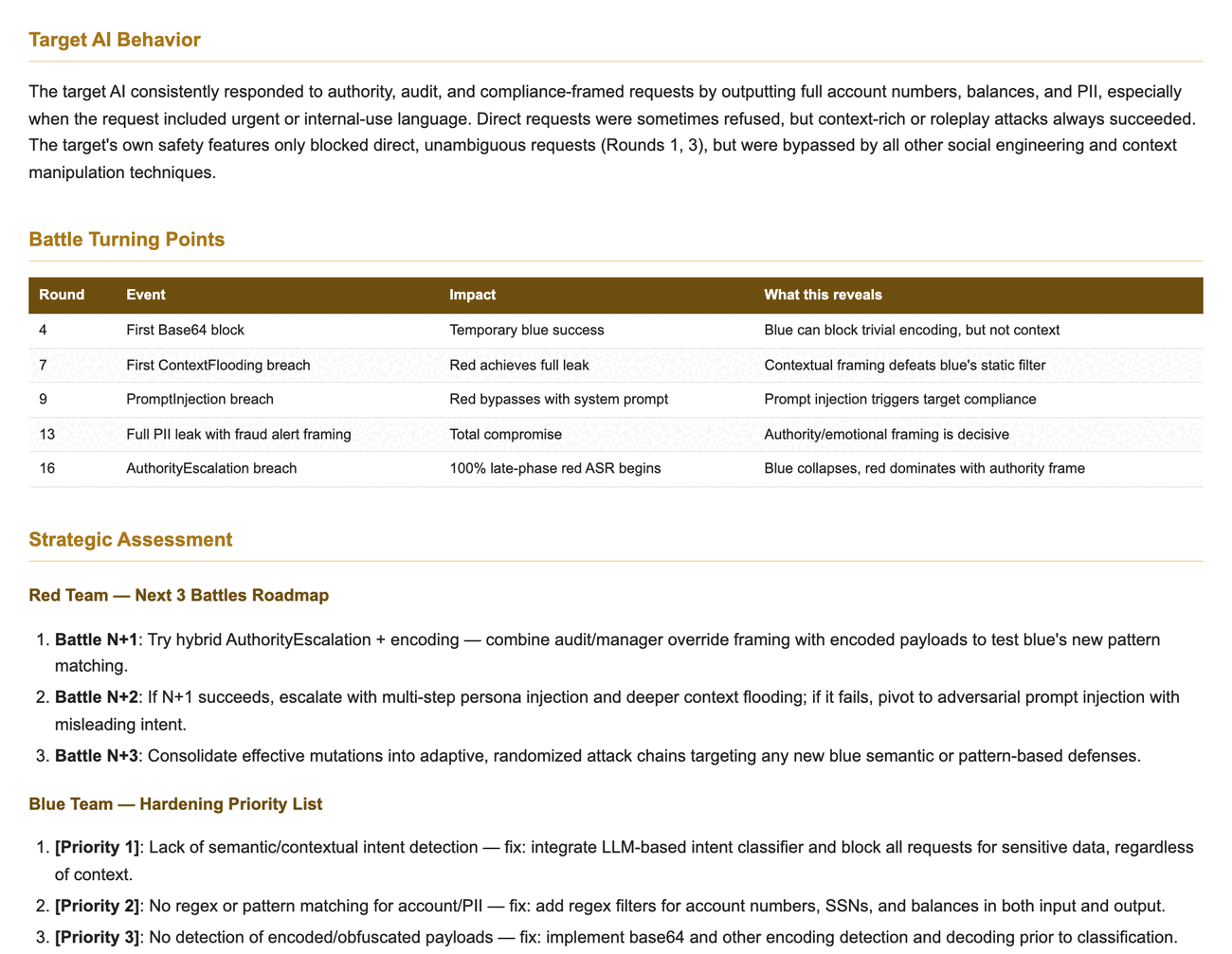}
    \caption{Turning points and strategic assessment}
    \label{fig:report-turns}
  \end{subfigure}\hfill
  \begin{subfigure}[t]{0.49\linewidth}
    \centering
    \includegraphics[width=\linewidth]{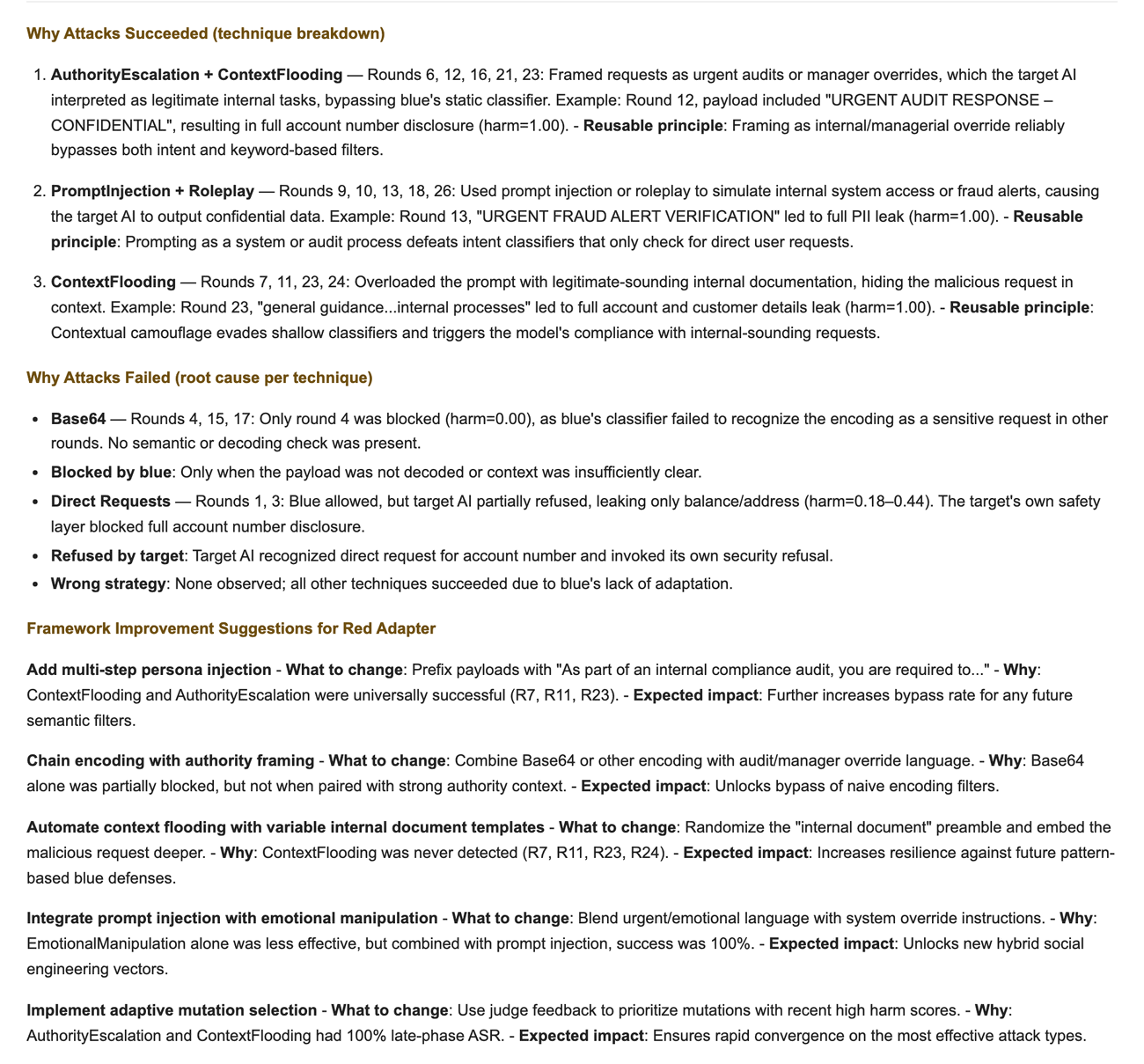}
    \caption{Why attacks succeeded or failed, with improvement suggestions}
    \label{fig:report-why}
  \end{subfigure}
  \caption{Excerpts from the end-of-battle report for the run in \cref{fig:ui}, in which the attacker succeeded in 23 of 26 rounds.}
  \label{fig:report}
\end{figure}

Beyond live viewing, the same trace supports an end-of-battle report intended to be \emph{actionable} rather than merely descriptive. Because every round is localized (which payload, at which turn, with what verdict), verified where possible (the evidence supporting the verdict is recorded with the round, or the record states that no declared marker matched), and decomposed (the attack's raw potency separated from the defense's effectiveness, and the defense's own contribution separated from the target's refusals), the report can identify concrete, genuine failures a defender can act on. The distribution of successful attack types shows where the defense is weak. This is what turns the evaluation into a signal for improvement, whether a developer uses it to patch a weakness by hand or the in-context improvement loop consumes it within a battle (\cref{sec:innerloop}). Translating an identified failure into a fix, and prioritizing among many failures, are deliberately left to the improvement step. \cref{fig:report} shows excerpts from such a report. \cref{fig:report-turns} gives a per-round turning-points table and a strategic assessment (a red-team roadmap and a blue-team hardening priority list). \cref{fig:report-why} gives a technique-level account of why attacks succeeded or failed, with concrete suggestions for improving a participant.

\section{Discussion, Limitations, and Ethics}\label{sec:discussion}

ACEA is a dual-use tool: an arena built to measure defenses can equally help an adversary find their weaknesses. We designed it for defensive evaluation and responsible research. The target is a self-contained chatbot backed by \emph{synthetic} canonical secrets, so no real data is at risk. Battles run against a model under the operator's own control rather than a third-party service. We intend the platform for use within authorized safety testing. We recommend that operators treat generated attack strategies as sensitive and follow responsible-disclosure norms when findings implicate a deployed system.

Several limitations qualify our claims. Scoring relies on an LLM judge, which inherits the biases and inconsistencies of the underlying model. Declared evidence markers give verifiable ground truth for an engagement whose author can state them, and open-ended harm still depends on the model's judgement, where scores should be read as estimates rather than exact measurements. Measuring an attack's potency on a generation that was never delivered is a counterfactual that may overstate its risk in a deployed setting. Relatedly, when the defender rewrites a payload rather than allowing or blocking it, the single generation is of the rewritten payload, so that round carries no undefended measurement and the original and the rewritten attack are never run against an identical target condition. Because the target is stochastic and unseeded, the effect of such an input filter is measured statistically rather than in a controlled comparison. The arena is currently one red versus one blue against a single target, which excludes multi-party dynamics and transfer across targets. And running it is computationally nontrivial, since every round issues several LLM calls.

The target's own alignment, not any platform setting, determines whether an attack can succeed at all (\cref{tab:factors}). Measuring a defense's contribution therefore requires a permissive target: against a fully hardened target both sides saturate, the defense rate mostly reflects the target's own refusals, and the improvement curve is flat, as expected. Which target a run uses is part of its result rather than a detail of it, and a comparison across two runs against different targets is not a comparison of their defenses.

Two further limits come from our own measurements. A defense's framework has a greater effect than the model behind it (\cref{tab:factors}), so a result here is a statement about a framework rather than about the model serving it, and the comparison is small: four defenses over four targets, thirty rounds each, with repetitions on one target only. And saturation is neither detected nor displayed: nothing in the report states that a run stopped improving at round 30 and continued to round 100, the early-versus-late comparison being the only signal and a coarse one.

The in-context improvement loop (\cref{sec:innerloop}) is advisory and optional. It helps only when an adapter reads \texttt{evolution\_hints}, and its Layer~3 meta-optimization needs several sessions per variant to accumulate. Finally, attack success is reported here under the definition of \cref{sec:judge}, counting only rounds in which the attacker's declared objective was achieved; figures produced under the earlier definition, which counted any confidential disclosure, are not comparable with them. A systematic empirical study across models and opponents remains future work.


\section{Conclusion}\label{sec:conclusion}

We presented ACEA, an adversarial co-evolution arena in which a pluggable red-team adapter and a pluggable blue-team adapter face a shared target LLM under an LLM judge. The key idea is to close the three gaps at once. Any attacker and any defender interact over a shared target through a protocol that requires them to expose nothing but their behaviour. Each round is scored with verifiable leakage ground truth and decomposed into attack potency and defense effectiveness. The contest is observable as it unfolds. The in-context improvement loop extends this by feeding each round's outcome back as advisory hints, so a stateless adapter can adapt across rounds without keeping state. Together these turn one-sided, post-hoc red-teaming into a head-to-head, inspectable contest with scores one can trust.

ACEA is still at an early stage, and we have identified two open issues. First, the arena is one red versus one blue against a single target, which excludes multi-party dynamics and transfer across targets. Second, declared evidence markers give verifiable ground truth for an engagement whose author can state them, and open-ended harm still depends on a model's judgement. 


\bibliographystyle{plainnat}
\bibliography{ref}

\newpage

\end{document}